\documentclass[
    ,final            
  ]
  {aipproc}

\layoutstyle{8x11single}
\usepackage{amsmath, amssymb}
\usepackage{graphicx}
\usepackage{isotope}
\usepackage{subfigure}
\makeatletter
\newcommand\setcaptype[1]{\def\@captype{#1}}
\makeatother

\newcommand{\ve}[1]{\ensuremath{\mathbf{#1}}}
\newcommand{\n}[1]{\ensuremath{|\mathbf{#1}|}}

\newcommand{\qrec}{\ensuremath{Q_{\textrm{rec}}^2}}

\newcommand{\sNC}{\ensuremath{\sigma^{\textrm{NC}}}}

\newcommand{\MB}{MiniBooNE}
\newcommand{\MBC}{MiniBooNE Collaboration}
\def\lsim{\lesssim}

\def\be{\begin{equation}}
\def\ee{\end{equation}}

\graphicspath{{plot/}}

\begin{document}

\title{Consistent analysis of neutral- and charged-current\\ (anti)neutrino scattering off carbon}


\author{Artur M. Ankowski}
{
  address={INFN and Department of Physics,``Sapienza'' Universit\`a di Roma, I-00185 Roma, Italy}
}

\begin{abstract}
Good understanding of the cross sections for (anti)neutrino scattering off nuclear targets in the few-GeV energy region is a~prerequisite for the correct interpretation of results of ongoing and planned oscillation experiments. To clarify a~possible source of disagreement between recent measurements of the cross sections on carbon, we analyze the available data within an approach based on the realistic spectral function of carbon, treating neutral-current elastic (NCE) and charged-current quasielastic (CCQE) processes on equal footing. We show that the axial mass from the shape analysis of the MiniBooNE data is in good agreement with the results reported by the BNL E734 and NOMAD Collaborations. However, the combined analysis of the NCE and CCQE data does not seem to support the contribution of multinucleon final states being large enough to explain the normalization of the MiniBooNE-reported cross sections.
\end{abstract}

\maketitle


\section{Introduction}
The correct interpretation of the outcome of oscillation experiments requires a~precise knowledge of the (anti)neu{\-}trino cross sections. This is the case even in those experiments in which the event yields in near and far detectors are compared, because, in general, flux and backgrounds do not scale in a~simple manner~\cite{ref:MiniB_CC}.

The description of nuclear effects in neutrino scattering is now generally regarded as one of the main sources of systematic uncertainties in oscillation experiments. In particular, profound understanding of charged-current quasielastic (CCQE) interaction with nucleons bound in the nucleus plays a~crucial role.

The carbon nucleus is of special importance, because the targets applied in neutrino detectors often involve carbon compounds, such as mineral oil, poly{\-}styrene, or organic scintillators. The total $\isotope[12][6]{C}(\nu_\mu,\mu^-)$ cross section as a~function of energy has recently been measured with the NOMAD~\cite{ref:NOMAD} and \MB{}~\cite{ref:MiniB_CC} experiments.

In NOMAD, studying neutrinos of energy down to 3~GeV, CCQE events with and without knocked-out proton detected have been analyzed separately and, after adjusting the description of final-state interactions, shown to yield consistent results. Thanks to the 45-GeV average energy of the deep-inelastic-scattering events, the normalization has been determined from the well-known total inclusive charged-current (CC) cross section and from the purely leptonic process of inverse muon decay. The observed reduction of the cross section due to nuclear effects is $\sim$4\%.

In \MB{}, all events without pions detected have been classified as CCQE and the total cross section is extracted for energy up to 2~GeV. The high statistics of CCQE events ($\sim$$10\times$ those in NOMAD) allowed for obtaining, for the first time, the double differential cross section. Surprisingly, the reported total cross section for carbon is \emph{higher} than that for free neutrons. The size of the effect is $\sim$5\% at a~few hundred MeV, increasing to $\sim$15\% at neutrino energy higher than $\sim$850~MeV.

As a~20\% uncertainty of the cross section would have an important impact on determination of oscillation parameters~\cite{ref:Davide_PLB}, the difference between the NOMAD and \MB{} results requires very careful theoretical analysis.

In this paper, we cover both NCE and CCQE processes, treating them in an identical manner and comparing the obtained cross sections to those measured by the Brookhaven National Laboratory Experiment 734 (BNL E734)~\cite{ref:BNL_E734_NC}, \MB{}~\cite{ref:MiniB_CC,ref:MiniB_NC}, and NOMAD~\cite{ref:NOMAD}. Our analysis is presented in detail in Ref.~\cite{ref:carbon}.

\section{Description of the approach}\label{sec:approach}
We assume that the process of neutrino-nucleus interaction involves a~single nucleon and the remaining $(A-1)$ nucleons act as a~spectator system. In this regime, called the impulse approximation (IA), the neutrino-nucleus cross section is obtained convoluting the elementary neutrino-nucleon cross section with the hole and particle spectral functions (SFs) of the nucleus.
The hole spectral function $P_\textrm{hole}(\ve p, E)$ is the probability distribution of removing a~nucleon of momentum $\ve p$ and leaving the residual nucleus with excitation $E$ in its rest frame. The particle spectral function $P_\textrm{part}(\ve{p'},\mathcal T')$ describes the propagation of a~nucleon of momentum $\ve{p'}$ and kinetic energy $\mathcal T'$.

Realistic hole spectral functions for various nuclei have been obtained by the authors of Ref.~\cite{ref:Omar_LDA} in the local density approximation (LDA), consistently combining the shell structure of the nucleus with the correlation contribution obtained from theoretical calculations for uniform nuclear matter at different densities~\cite{ref:Omar_LDA,ref:Omar_NM}. The carbon SF of Ref.~\cite{ref:Omar_LDA}, employed in this paper, has been extensively used in the analysis of electron scattering data in various kinematical setups. We use the particle SF obtained in the LDA scheme by convoluting the momentum distribution of nuclear matter at different densities with the density profile of the carbon nucleus~\cite{ref:ABF10}.

To account for multinucleon processes, in scattering on nucleons bound in the carbon nucleus, the effective axial mass $M_A=1.23$~GeV is applied. While this purely phenomenological method cannot be expected to be accurate in any kinematical regime, its validity is quantitatively supported by the results of Nieves {\it et al.}~\cite{ref:Nieves_PLB} for the double differential cross section in a~broad kinematical range of \MB{}. Moreover, in the context of our work, the kinematical setup of NOMAD does not differ significantly from that of \MB{}. It is a~simple consequence of the fact that (quasi)elastic processes constrain the high-$\n q$ contribution to appear solely at high $Q^2$, making it negligibly small due to the nucleon form factors. Hence, the phenomenological approach to multinucleon processes seems to be applicable also in comparisons to the NOMAD results.

In this paper, we consider the total cross sections and the flux-averaged differential cross sections $d\sigma/dQ^2$ for neutrino and antineutrino (quasi)elastic scattering. As FSI may cause only a~redistribution and a~shift of the strength, they do not affect the total cross section. The differential cross sections $d\sigma/dQ^2$ are expected to be modified by FSI at $Q^2 \lsim 0.15$ GeV$^2$~\cite{ref:Omar_oxygen}, because only at this kinematics the real part of the optical potential significantly changes the typical energy of the knocked-out nucleon. However, in the low-$Q^2$ regime, the validity of the impulse approximation, underlying our calculations, becomes questionable~\cite{ref:ABF10}. Therefore, in this article, FSI are not taken into account.

\section{Comparison to the NCE and CCQE data}
From the neutrino experiments which obtained results on NCE scattering off nuclei, the highest event statistics to date have been collected by the BNL E734 and \MB{} Collaborations.

In the BNL E734 experiment~\cite{ref:BNL_E734_NC}, studying $\nu p$ and $\bar\nu p$ NCE interactions, the target was composed in 79\% of protons bound in carbon and aluminium and in 21\% of free protons. The mean value of neutrino (antineutrino) energy was 1.3~GeV (1.2~GeV). Determination of the $\nu$ and $\bar\nu$ fluxes involved fitting to the CCQE event sample. Based on 1686 (1821) candidate $\nu p$ ($\bar\nu p$) events surviving the cuts, the flux-averaged differential cross sections ${d\sNC}/{dQ^2}$ for scattering off the BNL E734 target were extracted~\cite{ref:BNL_E734_NC}.

\begin{figure}
\setcaptype{figure}
\centering
    \subfigure[]
    {
    \includegraphics[width=0.4\textwidth]{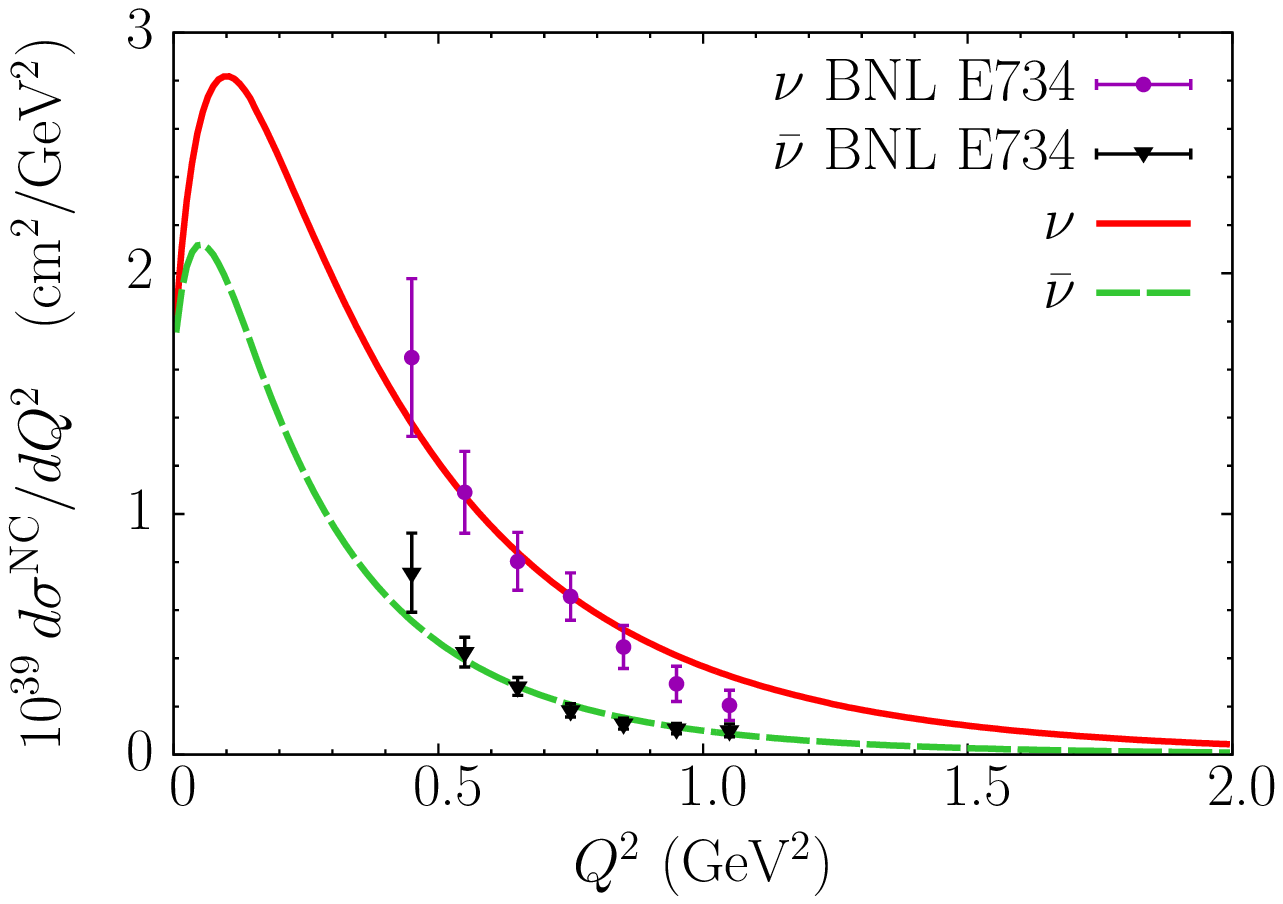}}
    \hspace{3em}
    \subfigure[]
    {\label{fig:MB_NC}
    \includegraphics[width=0.4\textwidth]{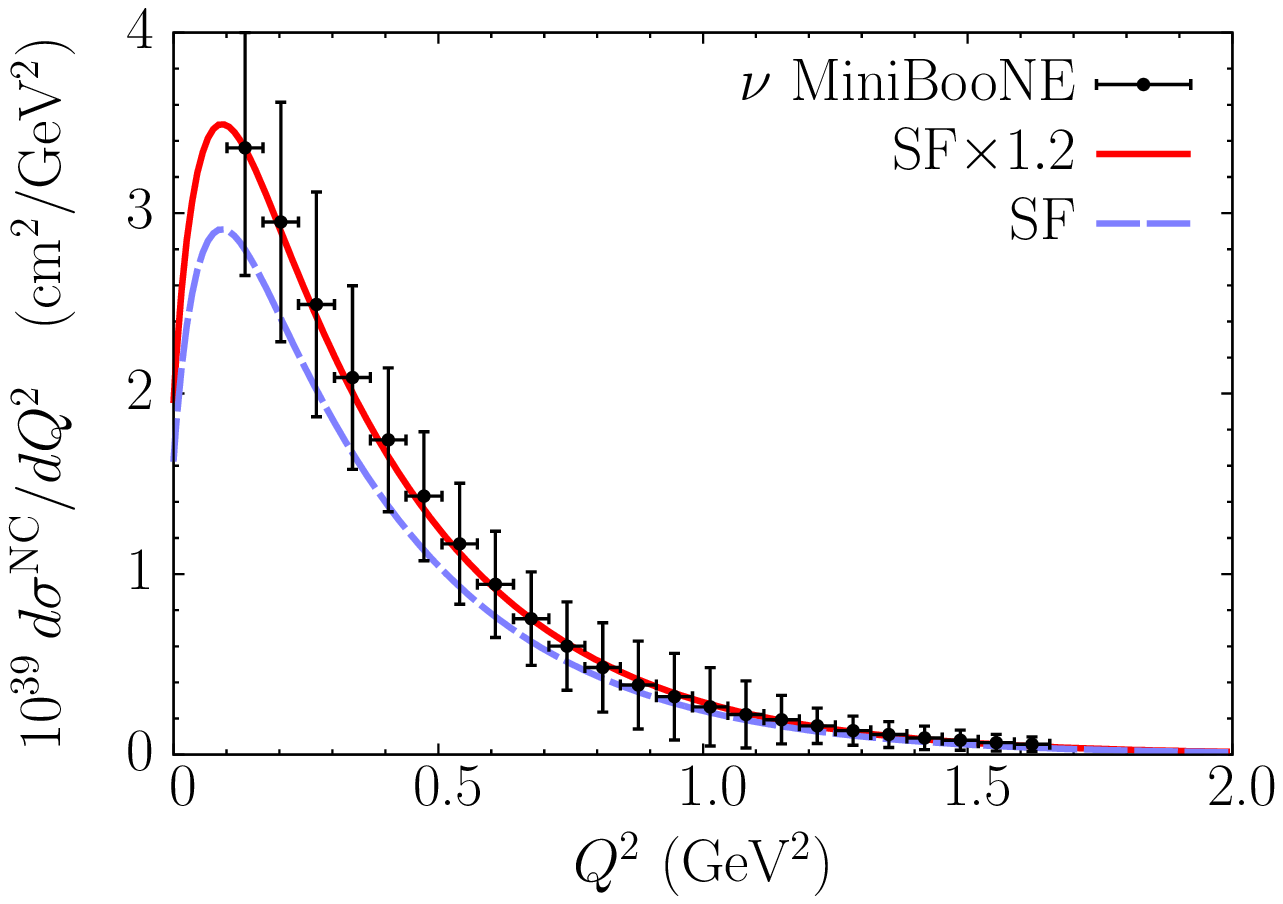}}
\caption{\label{fig:NC}Differential NCE cross sections $\sNC/dQ^2$ in (a) BNL E734 and (b) \MB{}. The SF calculations are compared with the data from Refs.~\cite{ref:BNL_E734_NC} and~\cite{ref:MiniB_NC}, respectively. The error bars show the statistical and systematic uncertainties added in quadrature. In panel (b), the theoretical result multiplied by 1.2 is also shown for comparison.
}
\end{figure}

Figure~\ref{fig:NC}(a) shows that the SF calculations provide a~fairly good description of the BNL E734 data~\cite{ref:BNL_E734_NC} in the whole range of $Q^2$. Note that the agreement seems to be better with the
higher-statistics antineutrino data. This is also the case for neutrinos in the region of the lowest uncertainty, $0.5\leq
Q^2\leq0.8$~GeV$^2$.

The \MB{} experiment, using the Cherenkov detector filled with mineral oil (CH$_2$), has been sensitive to both $\nu p$ and $\nu n$ NCE scattering~\cite{ref:MiniB_NC}. In neutrino mode, its beam has been composed almost exclusively of muon neutrinos, with an average energy of 788~MeV. The neutrino flux at the detector has been determined by a~Monte Carlo simulation~\cite{ref:MiniB_flux}. The accuracy of the shape determination has been proven by the comparison of the observed and predicted $\nu_\mu$ energy distribution in the CCQE event sample. However, the normalization of the measured distribution of CCQE events is reported to be higher by a~factor of $1.21\pm0.24$ than the calculated one~\cite{ref:MiniB_flux}. The \MBC{} recorded in neutrino mode 94531 candidate events surviving the NCE cuts, which allowed extraction of the differential cross section with unprecedented precision.

Figure~\ref{fig:NC}(b) shows that our calculations reproduce the \emph{shape} of the \MB{} NCE $\nu$ cross section very accurately, both in the region of lower $Q^2$ and in the tail. However, to match the absolute scale of the \MB{} data, it was necessary to multiply the SF results by a~factor of 1.2, consistent with the normalization discrepancy of $1.21\pm0.24$ observed by the \MBC{} in the CCQE analysis~\cite{ref:MiniB_kappa}.

\begin{figure}
\setcaptype{figure}
\centering
    \subfigure[]
    {
    \includegraphics[width=0.4\textwidth]{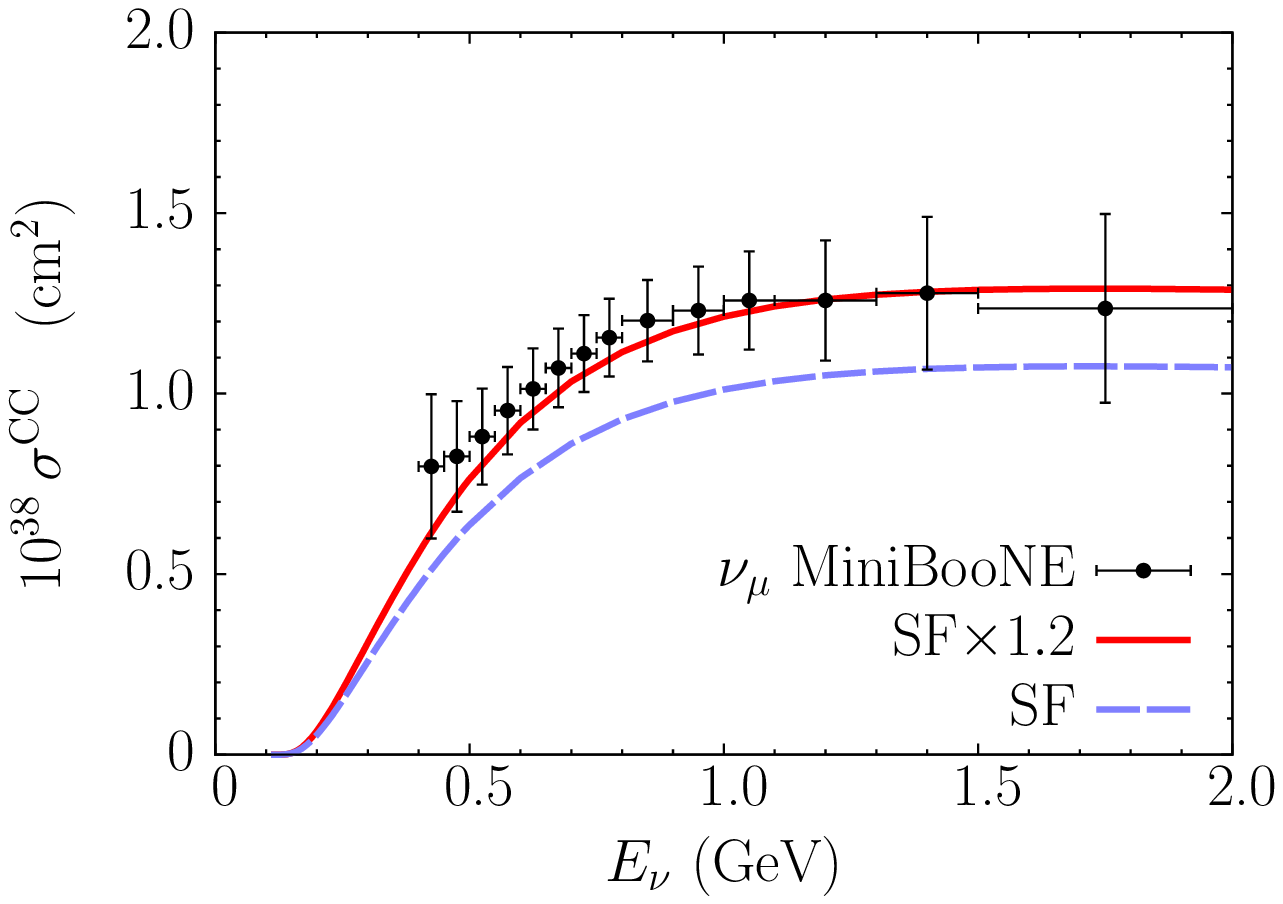}}
    \hspace{3em}
    \subfigure[]
    {
    \includegraphics[width=0.4\textwidth]{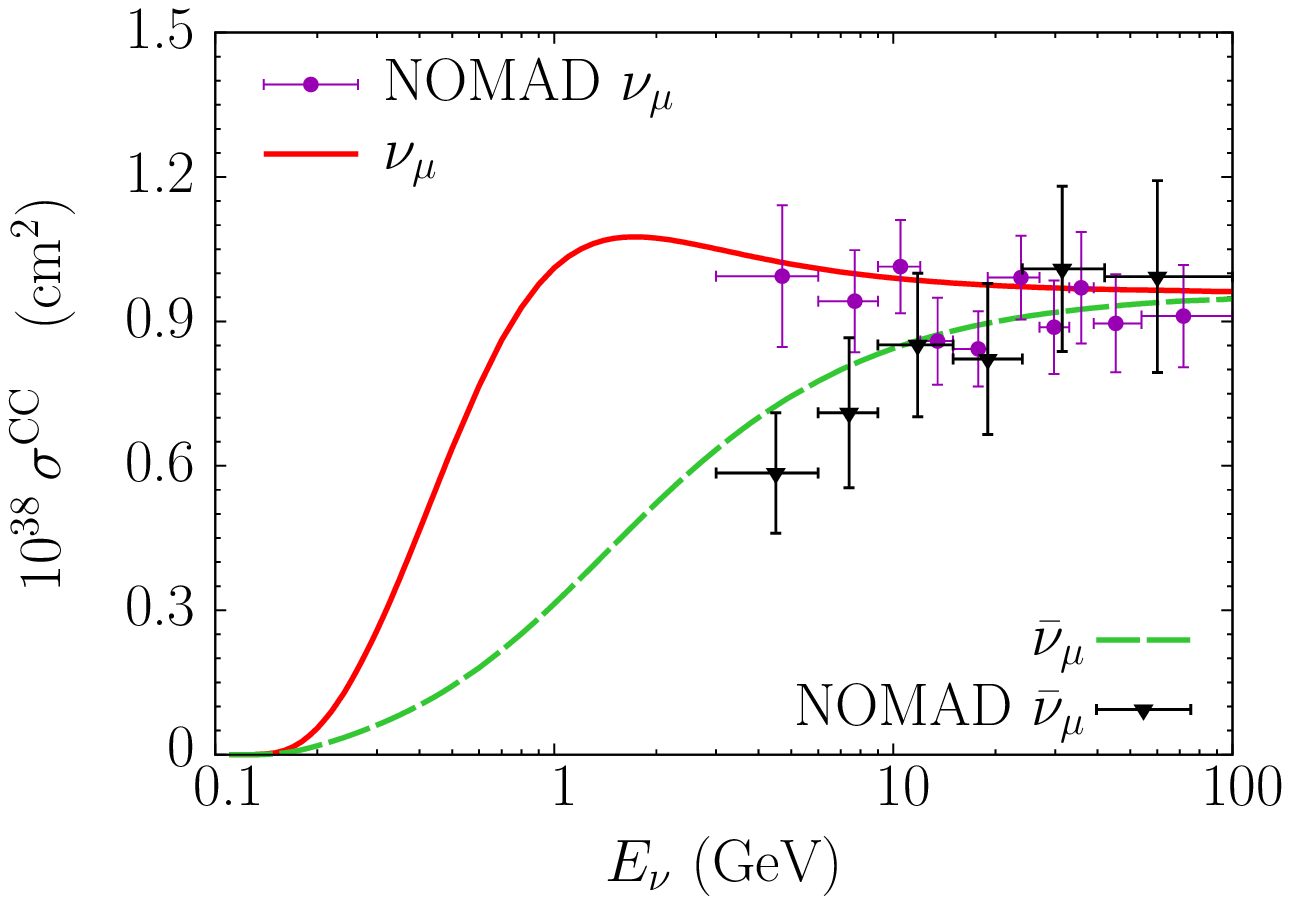}}
\caption{\label{fig:CC}Total CCQE cross sections for muon (anti)neutrino scattering off carbon. The SF results are compared with the data extracted from the (a) \MB{} and (b) NOMAD experiments. The error bars show the total uncertainties. In panel (a), the calculation multiplied by 1.2 is also shown for comparison.
}
\end{figure}

To clarify a~possible source of the normalization discrepancy, it is useful to check how our approach describes the CCQE data.
From the collected in neutrino mode 146070 events passing the cut, the \MBC{} has extracted, among others, the flux-unfolded total CCQE cross section~\cite{ref:MiniB_CC}. As presented in Fig.~\ref{fig:CC}(a), the energy dependence of our result is in good agreement with the experimental points, however, also in this case the factor of 1.2 is required to match the normalization.

From our analysis, it consistently emerges that, while correctly describing the shape of the \MB{}-reported cross sections, our calculations fail to reproduce their absolute scale, remaining inadequate by 20\%.

In this context, it is interesting to compare our results to the total $\isotope[12][6]{C}(\nu_\mu,\mu^-)$ and $\isotope[12][6]{C}(\bar\nu_\mu,\mu^+)$ cross sections measured with the recent NOMAD experiment~\cite{ref:NOMAD}. Due to the high energy of its broad-band neutrino (antineutrino) beam with the mean value of 25.9~GeV (17.6~GeV), the normalization was precisely determined from a~large sample of the deep-inelastic-scattering and inverse-muon-decay events. Using a~drift-chamber detector, a~total of 14021 (10358 single track and 3663 double track) neutrino and 2237 antineutrino events surviving the CCQE cuts were recorded.

In NOMAD, the neutrino events with more than two tracks (the charged lepton plus one proton with momentum $\geq300$ MeV) detected would have been removed from the CCQE event sample.
On the other hand, in \MB{}, all events with no pions detected, including two (or more) nucleon knockout events, contribute to the CCQE cross section.

Additional nucleons may originate from interactions between the struck nucleon and the spectator system in the final state, from the nucleon-nucleon correlations
in the initial state, and from reaction mechanisms other than nucleon knockout, such as those involving meson-exchange currents.

The effect of multiproton states due to FSI on the CCQE event sample was estimated in Monte Carlo simulations of the NOMAD experiment to be very small~\cite{ref:NOMAD}.
Because the nucleon-nucleon correlations produce in nuclei overwhelmingly proton-neutron pairs, they may have an influence on the track-based measurement of the CCQE cross section of neutrinos, but (in the absence of FSI) not on that of antineutrinos. Analyzing the nucleons with momentum larger than 300~MeV, we find that in the kinematics of by the NOMAD experiment, scattering off strongly correlated nucleon pairs yields $\sim$6\% of the inclusive cross sections. This number should be considered as an upper bound of a~difference between the NOMAD- and \MB{}-reported cross sections that the approach of this paper is able to describe.

Figure~\ref{fig:CC}(b) shows that our calculations of the total inclusive CCQE $\nu_\mu$ and $\bar\nu_\mu$ cross sections are in good agreement with the data obtained by the NOMAD Collaboration~\cite{ref:NOMAD}. We want to emphasize that their normalization is not scaled by additional factors. Although the SF results are higher by $\sim$6\% than the NOMAD best fit, this difference is less that the $\sim$8\% ($\sim$11\%) systematic uncertainty of neutrino (antineutrino) data. We observe that subtraction of the correlated contribution would bring the neutrino result in perfect agreement with the experimental points. However, this may be a~pure coincidence, as the analogical difference for antineutrinos cannot be explained in the same manner. Additionally, the results of Ref.~\cite{ref:MiniB_kappa} suggest that the approach of this paper may, to some extent, overestimate the cross section, owing to inappropriate description of the low-$Q^2$ contribution.

From the total CCQE $\nu_\mu$ cross section, the NOMAD Collaboration extracted the axial mass $1.05\pm0.02\textrm{(stat)}\pm0.06\textrm{(syst)}$ GeV. We estimate that in the SF approach that would correspond to 1.17~GeV, due to stronger quenching of the cross section. This value has been obtained without subtracting the correlated strength and, as such, gives the lowest $M_A$ required to fit the NOMAD data using our approach.

Analyzing the shape of the $\qrec$ distribution of double-track CCQE $\nu_\mu$ events for $0.2\leq \qrec\leq 4\textrm{ GeV}^2$, the NOMAD Collaboration found $M_A=1.07\pm0.06\textrm{(stat)}\pm0.07\textrm{(syst)}$ GeV. Within the quoted uncertainties, this value is marginally consistent with our estimate based on the total cross section.

\section{Discussion}\label{sec:Discussion}
The axial form factor $F_A(Q^2)$ is typically parametrized in the dipole form. Its value at \mbox{$Q^2=0$} is known from the neutron beta-decay measurements, whereas the dependence on $Q^2$ is governed by the axial mass $M_A$.

Because of the sizable contribution of the axial form factor to the (anti)neutrino cross section, the axial mass may be determined from the total cross sections or from shape fit to the event distribution with respect to some kinematic variable, e.g., $Q^2$. For free nucleons these two methods are equivalent, provided the dipole parametrization holds true. For bound nucleons, however, this may no longer be the case, due to inaccuracies of the applied description of nuclear effects. Note that different approaches could yield nearly identical (single) differential cross sections and, at the same time, different values of the total cross section, or {\it vice versa}.

We made use of this observation, showing for CCQE interaction that, within the SF approach, the axial mass extracted by the \MBC{} from the first shape analysis of the $\qrec$ event distribution~\cite{ref:MiniB_kappa} is in good agreement with the total $\isotope[12][6]{C}(\nu_\mu,\mu^-)$ cross section measured with the NOMAD experiment~\cite{ref:NOMAD}.

In Ref.~\cite{ref:MiniB_kappa}, the \MBC{} interpreted the higher value of the axial mass, extracted from the shape of the $\qrec$ distribution of CCQE(-like) $\nu_\mu$ events, as an effective method of accounting for the nuclear effects neglected in the RFG model. In fact, the processes which cannot be accounted for within the IA framework have recently been shown to contribute to CCQE-like scattering, increasing the total cross section and \emph{decreasing} the slope of the $\qrec$ distribution of events. Nieves {\it et al.}~\cite{ref:Nieves_PLB} have observed that these effects may effectively be described using a~higher value of the axial mass, confirming the interpretation of the \MBC{}.

Nevertheless, the results of Nieves {\it et al.}~\cite{ref:Nieves_PLB} suggest that the neutrino flux in the \MB{} experiment is underestimated by $9.2\pm3.5$\% when $M_A=1.077$~GeV is used in the approach based on the local RFG model. Because the total cross sections obtained in the RFG model are typically higher by $\sim$10\% than those in the SF approach, the 20\% normalization discrepancy that we have observed in comparison to the \MB{} data seems to be in perfect agreement with the finding of Nieves {\it et al}~\cite{ref:Nieves_PLB}. Note that the \MB{} flux estimate is based on the extrapolation of the cross section for $\pi^\pm$ production in $p$-Be scattering measured on a~thin target to a~target 35 times thicker. Such calculations are known to involve severe difficulties~\cite{ref:HARP_thickTarget}. The idea that the \MB{} flux might have been underestimated is not new~\cite{ref:Kopp}. Should it turn out to be correct, this would reconcile the \MB{} and NOMAD data in a~simple way.

In various interaction channels, the ratio of the events recorded with the \MB{} experiment to those predicted in the Monte Carlo simulation exceeds unity. It is interesting to note that the size of the ratio is similar: $1.21\pm0.24$ in CCQE scattering~\cite{ref:MiniB_kappa}, 1.23 in CC charged-pion production~\cite{ref:MiniB_piC}, and $1.58\pm0.05\textrm{(stat)}\pm0.26\textrm{(syst)}$ in CC neutral-pion production~\cite{ref:MiniB_pi0}. The quoted figures refer to the simulations with $M_A=1.23$~GeV and the parameter~$\kappa$, governing the enhancement of the Pauli blocking effect, set to 1.019. We checked that our approach and the model applied by \MB{} yield the total CCQE cross section differing by $\leq2.7\%$ for the neutrino energy higher than 350~MeV. Therefore, we may conclude that our 20\% normalization discrepancy is in good agreement with the \MB{} simulations and does not seem to be limited to NCE and CCQE interactions.

The latter conclusion is supported by the measurement of the inclusive CC $\nu_\mu$ cross section with the SciBooNE experiment~\cite{ref:SciBooNE}, using the same neutrino beam as \MB{}. The obtained result is larger than its Monte Carlo estimate by a~factor of 1.12 ($\kappa=1.0$) or 1.29 ($\kappa=1.022$), depending on details of the simulation.

Our calculations underestimate the absolute values of the \MB{}-reported cross sections, while being in a~good agreement with their shape and with the NOMAD data. In the NOMAD analysis, CCQE $\nu$ events are defined as containing at most one proton detected. When the proton's kinetic energy is measured to exceed 47~MeV (momentum $\geq300$~MeV), the event is classified as a~double-track one. The cross section extracted from the single- and double-track events, composing 73.9 and 26.1\% of the collected sample, respectively, is shown not to differ. Therefore, only those two- and multinucleon final states (2NFS) which involve additional protons of kinetic energy lower than 47~MeV and any neutrons contribute to the NOMAD result. Moreover, the contributions of the 2NFS to the single- and double-track samples seem to be equal and do not show energy dependence.

In the \MB{} analysis, CCQE events are required exclusively not to involve pions. Hence, the cross sections may be increased by a~broader class of 2NFS, involving two or more protons, each of kinetic energy $>47$~MeV. As a~consequence, the 20\% discrepancy between our calculations and the \MB{} data could, in principle, be ascribable to more sizable contributions of 2NFS than accounted for in our approach by using the effective axial mass of 1.23~GeV. Lacking apparent dependence on energy, it is, however, constrained by the NOMAD results. Note that
\begin{itemize}
\item[(i)]{the discrepancy in the NCE channel stems from the final-state nucleons with $50\leq T\leq 650$~MeV, where $T$ is their kinetic energy in total, and appears to the same extent over the whole range of $T$;}
\item[(ii)]{because the contributions of 2NFS to the \MB{}-reported NCE and CCQE cross sections seem to be equal, the kinematics of the knocked-out nucleons in these two cases should not differ significantly; and}
\item[(iii)]{the NOMAD results constrain the missing strength to involve at least two protons with $T\geq94$ MeV.}
\end{itemize}

The discrepancy between our calculations and the NCE cross section measured with \MB{} is not limited to $T\geq94$~MeV, corresponding to $Q^2\geq0.177$ GeV$^2$ and remains constant at the interval $50\leq T\leq 650$~MeV, which suggests that all the  contributing 2NFS channels are open below $T=50$~MeV. These features do not seem to be consistent with the hypothesis that 2NFS contribute to the \MB{} and NOMAD data in a~different manner, but rather point to the flux uncertainty in \MB{} being higher than reported.

\section{Summary}\label{sec:Summary}
In this paper, we have applied the spectral function approach to describe nuclear effects in (anti)neutrino scattering off carbon nucleus, treating in a~consistent manner NCE and CCQE interactions. We have considered a~broad energy range, from a~few hundreds of MeV to 100~GeV. The dipole parametrization of the axial form factor with the cutoff mass 1.23~GeV has been used, as determined from the shape of the $\qrec$ distribution of CCQE $\nu_\mu$ events by the \MBC{} in Ref.~\cite{ref:MiniB_kappa}. This effective method of accounting for two- and multinucleon final-state contributions to the cross section seems to be justified in view of recent results of Nieves {\it et al.}~\cite{ref:Nieves_PLB}.

It has been shown that our approach provides a~fairly good description of the NCE $\nu$ and $\bar\nu$ differential cross sections $d\sigma/dQ^2$ measured with BNL E734. A good agreement has been found with the total CCQE $\nu_\mu$ and $\bar\nu_\mu$ cross sections from NOMAD. While our calculations provide very accurate description of the shape of the NCE neutrino differential cross section $d\sigma/dQ^2$ obtained from \MB{}, they underestimate its absolute value by 20\%. The same discrepancy is observed for the flux-unfolded total CCQE cross section reported by the \MBC{}.

The difference between the total $\isotope[12][6]{C}(\nu_\mu,\mu^-)$ cross section from NOMAD and \MB{} is sometimes attributed to a~sizable contribution of two- and multinucleon final states. This reasoning is based on the fact that in the latter experiment, nucleons knocked out from the nucleus in CCQE neutrino scattering have not been detected.

We have argued, however, that the NCE data provide an indication of the kinematics of nucleons in CCQE scattering, because the \MB{} results strongly constrain the allowed differences between nuclear effects in NCE and CCQE $\nu_\mu$ interactions. In the kinematic region used to extract the NCE cross section, we find no evidence for the contribution of two- and multinucleon final states which would not have contributed also to the CCQE cross sections reported by the NOMAD Collaboration. Therefore, the discrepancy between the results from the \MB{} and NOMAD experiments seems more likely to be ascribable to underestimated flux uncertainty in the \MB{} data analysis.

\begin{theacknowledgments}
The author would like to thank Omar Benhar for illuminating discussions and careful reading of the manu{\-}script. Thanks are also due to the Organizers of NuInt12 for providing partial financial support. This work was supported by the INFN under Grant No. MB31.
\end{theacknowledgments}

\bibliographystyle{aipproc}

\end{document}